\documentclass{article}
\usepackage{frascatiphys}
\usepackage{graphicx}
\def\ls{{_<\atop^{\sim}}}
\def\gs{{_>\atop^{\sim}}}
\begin{document}
\title{Confirming the Detection of two WHIM Systems along the Line of Sight to 1ES~1553+113}
\author{ F. Nicastro$^{1,2}$ \\
$^1$ {\em INAF - Osservatorio Astronomico di Roma}, \\ {\em Via Frascati, 33, 00078 Monte Porzio Catone (RM), Italy} \\
$^2$ {\em Harvard-Smithsonian Center for Astrophysics, Cambridge, MA, USA} }
\maketitle
\baselineskip=11.6pt

\begin{abstract}
We present a re-analysis, with newly acquired atomic data, of the two detections of two highly ionized intervening OVII absorbers reported by Nicastro and 
collaborators (2018). 
We confirm both intervening Warm-Hot Intergalactic Medium OVII detections, and revise statistical significance and physical parameters of the absorber at 
$z=0.4339$ in light of its partial contamination by Galactic interstellar medium NII K$\alpha$ absorption.  
\end{abstract}
\baselineskip=14pt

\section{Introduction}
Hydrodynamical simulations for the formation of structures in the Universe predict that, starting at redshift of $z\sim 2$, diffuse baryons in the intergalactic 
medium (IGM) condense into a filamentary web and undergo shocks that heat them up to temperatures T$\simeq 10^5 - 10^7$ K, making their by far largest 
constituent, hydrogen, mostly ionized (e.g. [1,2]). At the same time, galactic outflows powered by stellar and AGN feedback, enrich these baryons with metals (e.g. 
[2]). 
How far from galaxies these metals roam, depends on the energetics of these winds but it is expected that metals and galaxies will be spatially correlated.
This shock-heated, metal-enriched medium, known as Warm-Hot Intergalactic Medium (WHIM), is made up of three observationally distinct phases: 
(1) a warm phase, with T$\simeq 10^5 - 10^{5.7}$ K, where neutral hydrogen is still present with ion fraction $f_{HI} > 10^{-6}$ and the best observable metal ion 
tracers are OVI (with main transitions in the FUV) and CV (with transitions in the soft X-rays); 
(2) a hot phase with T$\simeq 10^{5.7} - 10^{6.3}$ K, where $f_{HI} \simeq 10^{-6} - 10^{-7}$ and OVII (with transitions in the soft X-rays) largely dominates metals 
with ion fractions near unity; 
and (3) an even hotter phase (T$\simeq 10^{6.3} - 10^7$ K), coinciding with the outskirts of massive virialized groups and clusters of galaxies, where HI and 
H-like metals are present only in traces (e.g. [1]).

The warm phase of the WHIM has indeed been detected and studied in detail in the past few years and is estimated to contain an additional 15\% fraction 
of the baryons (e.g. [3,4] and references therein). This brings the total detected fraction to 61\% but still leaves us with a large (39\%) fraction of elusive baryons, 
which, if theory is correct, should be searched for in the hotter phases of the WHIM. In particular, the diffuse phase at T$\simeq 10^{5.7} - 10^{6.3}$ K should contain 
the vast majority of the remaining WHIM baryons, and it is traced by OVII. Optimal signposts for this WHIM phase are then OVII He$\alpha$ absorption lines, 
which however are predicted to be relatively narrow (Doppler parameter $b(O) \simeq 20 - 46$ km s$^{-1}$), extremely shallow (rest-frame equivalent widths 
EW$\ls 10$ m\AA), and rare. 
Such lines are unresolved by current X-ray spectrometers and need a signal to noise ratio per resolution element SNRE$\gs 20$ in the continuum 
to be detected at a single-line statistical significance$\gs 3 \sigma$. 
This requires multi-million second exposures against the brightest possible targets available at sufficiently high redshift ($z \gs 0.3$).

In this contribution we first summarize the fidings from our recent discovery of two intervening OVII-bearing absorption systems along the line of sight to the 
blazar 1ES~1553+113, at redshifts $z =0.3551$ and $z=0.4339$ $^{[5]}$, then introduce a slight revision of our recently published results$^{[5]}$ in light of newly 
determined measurements of wavelengths and oscillator strengths of the NII K$\alpha$ complex (McLaughlin, private communication) that make our own Galaxy's 
ISM contamination likely for the OVII K$\alpha$ line of the system at $z=0.4339$, and finally discuss the implications of our finding. 

Throughout the paper uncertainties are quoted at 68\% significance, unless explicitly stated.

\section{Intervening WHIM Systems along the Line of Sight to 1ES~1553+113} 
The detections of two WHIM systems at $z=0.4339$ and $z=0.3551$ in the XMM-{\em Newton} RGS spectrum of 1ES~1553+113, have been presented by [5]. 
Here we briefly summarize their main findings. 

The 8-33 \AA\ RGS spectrum shows a number of narrow (unresolved) line-like negative features (Fig. 1 in Extended Data - ED, hereinafter - of [5]), eight of which are 
securely identifiable as Galactic absorption lines (marked and labeled in blue in Fig. 1 of ED of [5]). 
Two additional unresolved absorption lines are detected in both RGSs at combined single-line statistical significances of $4.1 - 4.7 \sigma$  and $3.7 - 4.2\sigma$ 
(Fig. 1 and Table 1 in ED of [5])
\footnote{here, and throughout the paper, we report a range of statistical significance, where the upper boundary is the actual measured single-line statistical 
significance, while the lower boundary is the measured significance conservatively corrected for observed systematics in the RGS spectrum (details in [5]).}
, at wavelengths where (1) no strong Galactic absorption is expected (but see \S 3 for likely NII K$\alpha$ ISM contamination for one of these two lines) and 
(2) neither of the two spectrometers is affected by instrumental features due to cool-pixels in the dispersing detector (Fig. 1). These are the lines identified by 
[5] as intervening WHIM OVII He$\alpha$ at $z=0.4339 \pm 0.0008$ and $z = 0.3551^{+0.0003}_{-0.0015}$ (Table 1 in ED of [5]). 
An additional lower significance ($1.7 - 2\sigma$) line is detected at a $\lambda = 26.69\pm 0.09$ \AA, and is identifiable as OVII He$\beta$ at 
a redshift consistent with $z=0.4339 \pm 0.0008$ (Table 1 in ED of [5] and Fig. 1, where the sizes of the arrows are proportional to the relative strengths of the lines
\footnote{in particular, the size of the arrows of low-ionization lines are 
relative to the strength of the NI K$\alpha$ transition, while those of the high-ionization lines are relative to the strength of the He$\alpha$ transition.}
). 

Here we confirm these identifications, but in light of new laboratory-experiment revised positions and oscillator strengths of the lines of the NII K$\alpha$ triplet 
(McLaughlin, private communication: see \S 3), we slightly revise the physical parameters of the $z=0.4339$ WHIM system (and thus the implied WHIM OVII 
cosmological mass density estimate) and its statistical significance (see \S 3). 

\begin{figure}[htb]
    \begin{center}
        {\includegraphics[scale=0.5]{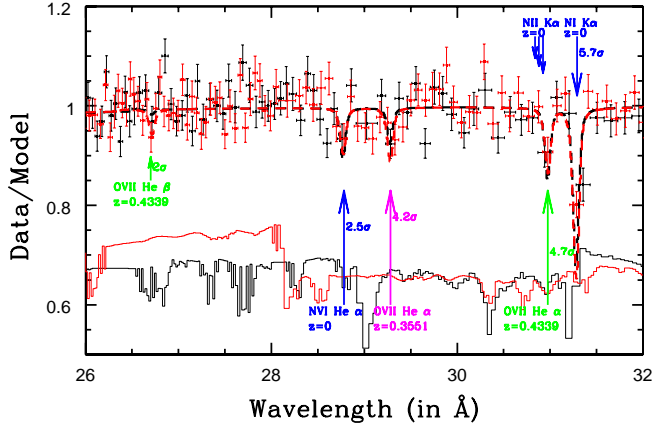}}
        \vspace{-3cm} \caption{\it Normalized raw RGS1 (black points) and RGS2 (red points) data of the blazar 1ES 1553+113, in the wavelength interval 
$\lambda = 6 - 32$ \AA. Thick dashed curves are RGS1 (black) and RGS2 (red) best-fitting model folded through the RGSs’ response functions. Thin solid curves 
at the bottom of the graph are RGS1 (black) and RGS2 (red) effective areas (in arbitrary units), showing instrumental features due to cool-pixels in the 
dispersing detectors.}
\label{fig1}
    \end{center}
\end{figure}

\section{Galactic NII Contamination for the $z=0.4339$ WHIM System}
In Fig. 1 the relatively strong line present in the data at a centroid $\lambda = 30.975 \pm 0.017$ \AA, is $\ge 50$ m\AA\ ($\ge$ one RGS resolution element) 
inconsitent with the theoretical (i.e. computed with the Hebrew University Lawrence Livermore Atomic Code - HULLAC$^{[6]}$ ; E. Behar, private communication) 
rest-frame wavelengths of the NII K$\alpha$ triplet (three blue arrows at $\lambda = 30.836$, 30.879 and 30.924 \AA) available to us at the time of publication of [5]. 
For this reason, this line was safely identified by [5] as the OVII He$\alpha$ transition imprinted by an intervening WHIM system $z=0.4339$. 

Laboratory positions and strengths of the three main NII K$\alpha$ lines, were already available in 2011$^{[7]}$, but were the outcome of the first early experiments 
done at the Optimized Light Source of Intermediate Energy laboratory (SOLEIL
\footnote{https://www.synchrotron-soleil.fr/en/about-us/what-soleil/soleil-3-questions}
) when the instability of synchrotron beam profiles was still poorly understood. 
New measurements for wavelengths and cross-sections of the NII ion have recently been perfomed at SOLEIL and results from the analysis of these new data have  
been made available to us (McLaughlin, private communication) and are shown (graphically) in Fig. 2. 
Fig. 2 shows two narrow portions of the RGS spectrum of 1ES~1553+113, $\lambda = 30.5-31.5$ \AA\ (top panel) and $\lambda = 23-24$ \AA\ (bottom panel). 
These are the spectral regions were the K$\alpha$ transitions of NI and NII (top panel) and OI and OII (bottom panel), lie. The arrows in Fig. 2 mark the positions 
of these lines and, as for Fig. 1, their relative size is proportional to the relative strengths of the transitions. 
The new laboratory measurements of the NII K$\alpha$ triplet indicate that the centroids of these lines are now consistent with the $\lambda = 30.975 
\pm 0.017$ line present in the data (Fig. 2, top panel). 

The question thus arises: can this line be entirely due to Galactic ISM absorption? 
A first problem with this hypothesis is that the centroids of the three NII K$\alpha$ lines are about 1 RGS resolution element apart from each other and the lines 
have different strengths. Thus, Galaxy's ISM NII K$\alpha$ absorption should imprint a relatively shallow, broad and skewed profile trough in the data, rather 
than the unresolved, symmetric, line-like feature present in the data. 

To test this possibility further, we used our {\em galabs} model$^{[8]}$ to self-consistently model the cold and mildly ionized ISM absoprtion components 
of our Galaxy along the line of sight to 1ES~1553+113

\subsection{Modeling the Cold-Neutral and Warm-Ionized ISM components in the RGS spectra of 1ES~1553+113}
The insterstellar medium of our Galaxy contains both Cold-Neutral and warm-ionized Metal-rich Medium (CNMM and WIMM$^{[8]}$) which attenuates the soft 
X-ray spectrum of both Galactic X-ray binaries and AGNs. The line of sight to 1ES~1553+113 is no exception. The RGS spectrum of 1ES~1553+113 clearly shows 
metal photo-electric absorption by neutral and mildly ionized oxygen and nitrogen (Fig. 2). 

We model both the bound-free (flattening of the long-wavelength X-ray powerlaw) and bound-bound (K$\alpha$ resonant lines from neutral metal ions) photo-electric 
absorption by the CNMM with a Tuebingen-Boulder ISM absorption component ({\em tbabs} in XSPEC), with solar abundances set to [9] and lower boundary of the 
hydrogen column density frozen to the weighted average measurement along this line of sight: $N_{HI} = 3.7 \times 10^{20}$ cm$^{-2}$ $^{[10]}$. 
The best-fitting N$_H$ is pegged to its lower boundary and the model reproduces well both the broad-band attenuation of the continuum at low energies and the 
K$\alpha$ lines of OI and NI (Fig. 2, blue histogram). 

This CNMM component does not include absorption by non-neutral metal species and thus cannot model either the strong OII K$\alpha$ (Fig. 2, bottom panel) or the 
weaker NII K$\alpha$ (Fig. 2, top panel) triplets 
\footnote{In Fig. 2 the arrows, and their relative sizes, shows the latest laboratory-measurement positions and relative strengths 
of the OII K$\alpha$ (Bizau et al., 2015) and NII K$\alpha$ (McLaughlin, private communication) transitions.}
. We thus add a WIMM component$^{[8]}$ to our model, with relative abundances set to Solar-like$^{[9]}$ and absolute metallicity free to vary, and refit the data. 
The best-fitting WIMM component has typical physical parameters (T$\sim 3000-5000$ K; N$_H = 1.85 \pm 0.07 \times 10^{20}$ cm$^{-2}$) and metallicity 
($Z = (0.52 \pm 0.09) Z_{\odot}$) and, together with the CNMM component, model excellently the OI K$\alpha$ line, the OII K$\alpha$ triplet and the NI K$\alpha$ 
line in the data, but cures only modestly the narrow line-like absorption deficit seen near the NII K$\alpha$ triplet (Fig. 2, red histogram). 

To model this additional feature, we add an unresolved (FWHM frozen to 10 m\AA) negative Gaussian to our model and refit the data. 
The best-fitting Gaussian has centroid $\lambda = 30.975 \pm 0.010$ \AA\ and EW$=10 \pm 3$ m\AA\ (i.e. a single line significance of $2.9-3.3 \sigma$).  
We therefore confirm the identification of this line as an intervening WHIM OVII K$\alpha$ line at $z=0.4339$, as in N18. 
Our final best-fitting model is the green histogram of Fig. 2.

We note that the best-fitting profile of the unresolved NII K$\alpha$ absorption triplet (red histogram), is in all (i.e. in wavelenths, shape and strength) similar to 
the gentle curvature seen in the continuum folded through the RGS effective area (blue histogram). This is an effective area feature and has been introduced by a recent 
correction made by the RGS calibration team (J. Kaastra, private communication) based on the data of the calibration sources Mkn~421 and PKS~2155-304. 
We think, instead, that this relatively narrow curvature in the data of calibration sources has an astrophysical (not instrumental) origin and is due to the 
ubiquitous ISM NII K$\alpha$ absorption bound to be imprinted in the X-ray aspectra of any astrophyscal source. 
By adopting this correction, thus, we are conservteily underestimating the actual strength and statistical significance of the intervening WHIM OVII K$\alpha$ 
absorption line at $z=0.4339$. 
\begin{figure}[htb]
    \begin{center}
        {\includegraphics[scale=0.5]{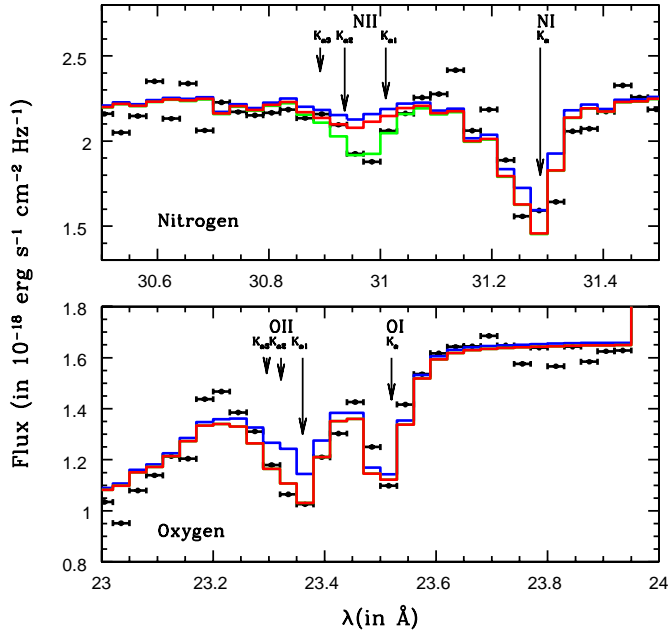}}
 \vspace{-1cm}       \caption{\it RGS spectrum of 1ES~1553+113 at $\lambda = 30.5-31.5$ \AA\ (top panel) and $\lambda = 23-24$ \AA\ (bottom panel). 
These are the spectral regions were the K$\alpha$ transitions of NI and NII (top panel) and OI and OII (bottom panel), lie.}
\label{fig2}
    \end{center}
\end{figure}

\section{Revised Comsological Mass Density of OVII K$\alpha$ WHIM Absorbers} 
The revised (compared to [5]) equivalent H column density and metallicity of the $z=0.4339$ WHIM system, are: 
N$_H = 0.7^{+0.5}_{-0.3} \times 10^{19}$ cm$^{-2} (Z/Z_{\odot})^{-1}$ and $0.05 < (Z/Z_{\odot}) < 0.2$ (see [5]). 
This gives an OVII WHIM cosmological mass density estimate $0.002 < \Omega_b^{WHIM} < 0.016$ (i.e. 9-70\% of the Universe's baryons). 

\section{Acknowledgements}
FN thanks B. McLaughlin for providing the new laboratory and theory data of the NII K$\alpha$ transitions. 
\section{References}
\end{document}